\documentclass[a4paper, 11pt]{article}
\date{May 2006}
\usepackage{epsfig}

\newcommand{\be}{\begin{equation}}
\newcommand{\ee}{\end{equation}}
\newcommand{\ba}{\begin{eqnarray}}
\newcommand{\ea}{\end{eqnarray}}
\newcommand{\bi}{\begin{itemize}}
\newcommand{\ei}{\end{itemize}}
\newcommand{\tr}{{\rm Tr\,}}

\newcommand{\<}{\langle} 
\renewcommand{\>}{\rangle}  
\newcommand{\eq}{Eq.~}

\newcommand{\fig}{Fig.~}
\newcommand{\tab}{Tab.~}
\newcommand{\la}{\label}

\newcommand{\Nfo}{\mathop{N_{\rm f}=0}}
\newcommand{\Nfii}{\mathop{N_{\rm f}=2}}

\newcommand{\tauint}{\mathop{\tau_{\rm int}}}

\newcommand{\half}{{\textstyle{\frac{1}{2}}}}

\newcommand{\mps}{\mathop{M_{\rm PS}}}
\newcommand{\Fps}{\mathop{F_{\rm PS}}}

\begin{document}
\begin{titlepage}

\begin{flushright}
DESY-06-080\\
HU-EP-06/14\\
SFB/CPP-06-26
\end{flushright}
\begin{centering}
\vfill

{\Large{\bf Exploring the HMC trajectory-length dependence \\
    of autocorrelation times in lattice QCD}}

\vspace{3cm}
\centerline{\psfig{file=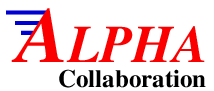,angle=0,width=3cm}}
\vspace*{0.5cm}

{\bf Harvey Meyer, Hubert Simma, Rainer Sommer}\\
\centerline{Deutsches Elektronen Synchrotron DESY}
\centerline{Platanenallee 6}
\centerline{D-15738 Zeuthen}

\vspace*{0.5cm}
{\bf Michele Della Morte, Oliver Witzel, Ulli Wolff}\\
\centerline{Humboldt-Universit\"at, Institut f\"ur Physik}
\centerline{Newtonstrasse 15}
\centerline{D-12489 Berlin}

\vspace*{2.0cm}

{\bf Abstract\\}
\end{centering}
\noindent 
We study autocorrelation times of physical observables
in lattice QCD as a function of the molecular dynamics 
trajectory length in the hybrid Monte-Carlo algorithm.
In an interval of trajectory lengths where energy and 
reversibility violations can be kept under control, 
we find a variation of the integrated autocorrelation times by 
a factor of about two in the quantities of interest.
Trajectories longer than conventionally used are found to
be superior both in the $\Nfo$ and $\Nfii$ examples considered here.
We also provide evidence that they lead to faster thermalization 
of systems with light quarks.
\vfill
\end{titlepage}

\setcounter{footnote}{0}
\section{Introduction\la{sec:intro}}
Hybrid Monte-Carlo (HMC)~\cite{Duane:1987de}
is the most widely used algorithm to 
simulate non-Abelian gauge theories in the presence of $N_f>0$ quark
flavours. For Wilson-type quark actions it was found,
in its most straightforward application,
to become prohibitively expensive in the 
regime of phenomenological interest (with two very light quarks)~\cite{Ukawa:2002pc}.
Recent advances~\cite{Luscher:2005rx,Hasenbusch:2001ne,AliKhan:2003br,Clark:2005sq}
however have led to a substantial lowering 
of the CPU time of simulations.
These more sophisticated forms of HMC typically 
have a number of parameters that have to be tuned to guarantee
an efficient simulation. One parameter that is however common
to all HMC algorithms is the length $\tau$ of the trajectories of `molecular 
dynamics' evolution, at the end of which an accept/reject step
is performed to correct for any finite-step-size errors.

The algorithm efficiency's dependence on $\tau$
has only been rarely and partially studied in the 
literature~\cite{DellaMorte:2003jj,Gehrmann:1999wr,Kennedy:1990bv,Mackenzie:1989us}.
The reason for this is easy to understand. If we neglect the cost 
of evaluating the Hamiltonian, performing
trajectories of length $\tau=2$ represents the same computational
cost as  twice as many trajectories of length $1$:
indeed, as discussed in the appendix, the average energy violations
along a trajectory at fixed step-size are weakly dependent on 
the length within a reasonable range, 
and the reversibility violations increase rather slowly
(see however~\cite{Liu:1997fs}). Hence, with a high acceptance rate 
in both cases,  to discriminate between
these two running modes one has to determine the autocorrelation times
of the observables of interest with some accuracy and reliability.
This in turn requires very high statistics, which one is normally not 
ready to invest into parameter optimization. 
In this letter we study this question in various situations where 
high statistics can be achieved, and  yet some realistic features
of difficult simulations are present. 

Whether in the end our conclusions carry over to other forms 
of HMC algorithms will require direct testing. However we show
that the dependence of the algorithm's efficiency on $\tau$ can be 
significant, and a factor around two in computing time can
easily be wasted in the case of an unfortunate choice of 
trajectory length.

\section{Data\la{sec:data}}
We employ the hybrid Monte-Carlo algorithm~\cite{Duane:1987de}.
The Hamiltonian governing the evolution of the molecular dynamics is 
\be
H = \half\sum_{\mu,x} \tr\{\Pi_\mu(x) \Pi_\mu(x)\} + S[U].
\la{eq:Hmd}
\ee
where the momenta $\Pi_\mu$ are traceless Hermitian matrices.
The gauge fields are thus updated according to  
$U^{\prime}_{\mu}(x) = \exp[i\Pi_{\mu}(x)d\tau]U_{\mu}(x)$.
In the $\Nfii$ simulations,
two pseudofermions are used to stochastically 
represent the determinant of the
O($a$) improved Wilson fermions: the fermionic part of $S$ 
results from even-odd-~\cite{Jansen:1996yt} and 
mass-preconditioning~\cite{Hasenbusch:2001ne} the 
Hermitian Dirac operator.
The gauge part of $S$ is the Wilson plaquette action.

When $\Nfii$, we use the Sexton-Weingarten integration 
scheme~\cite{Sexton:1992nu}, 
with a step-size four times larger for the fermionic forces;
the quoted step-size $d\tau$ always denotes the largest one.
A complete update cycle with trajectory length $\tau$
performs an integer number  $\tau/d\tau$ of such steps
followed by the Metropolis decision.
The system has then advanced by $\tau$
molecular dynamics (MD) units, and in the rest of this paper
all quantities referring to Monte Carlo time are given in MD units.
If for instance successive
measurements of observables are separated by $M$ trajectories,
an autocorrelation function $\Gamma(i)$ 
arises where $i$ refers to {\em successive measurements}.
From it we define the integrated
autocorrelation time, that is directly relevant for the statistical error,
in our MD units as
\be
\tauint = M\tau {\textstyle\left[\frac{1}{2} + \sum_{i \ge 1} \rho(i)\right]},\qquad
\rho(i)=\Gamma(i)/\Gamma(0).
\ee
In numerical estimates the sum has to be truncated. If we specify a window
$W$
(in MD units) this amounts to the restriction $i \le W/(M \tau)$. 
We  refer the reader to~\cite{Wolff:2003sm} for the definition of $\Gamma(i)$,
in particular for derived quantities, such as an effective mass.
The error bars shown in plots of $\rho$ are computed using 
\eq (E.11) of reference~\cite{Luscher:2005rx}.

We investigate the trajectory length dependence of 
autocorrelation times in three different systems (A,B,C: see \tab\ref{tab:param}).
All simulations were done in 32-bit arithmetic, except for system C (64-bit). 
The system D will be used as a playground for thermalization.
The lattice spacing is known in units of $r_0$~\cite{Sommer:1993ce}
at $1\%$ level for the $N_f=0$ runs and $5\%$ for the $\Nfii$ runs.
All lattices have Schr\"odinger functional boundary conditions.
Since we will be using somewhat longer trajectories than is usual, 
the question of reversibility violations arises. We check for this 
in the standard way (see e.g.~\cite{DellaMorte:2003jj}) by monitoring
the Hamiltonian variation $\<|\delta H|_\leftrightarrow\>$ 
under the following operations : a trajectory `forward', 
reversing the sign of the momenta, and integrating back to the starting point.

\begin{table}
\begin{center}
\begin{tabular}{|c|c|c|c|c|c|c|c|c|}
\hline
         & $L^3\times T$ &  BF & $N_{\rm f}$  & $1/d\tau$  & $\beta$ &  $\kappa$  &  
                $a/r_0$ [ref]& $a\mps$\\
\hline
A        &  $8^4$ & $A$ & 0 & 30 &    $6.086$    & - & 0.16~\cite{Necco:2001xg}  &  - \\
         &  $12^4$ &$A$ & 0 & 30 &    $6.364$ & - &0.11~\cite{Necco:2001xg}   &  - \\
         &  $16^4$ &$A$ & 0 & 30 &    $6.57$ & - &0.08~\cite{Necco:2001xg}   &  - \\
B        & $8^3\times32$  & 0& 0 &  50   &  6.0   &   0.1338 & 0.19~\cite{Necco:2001xg} & 0.38(2) \\
C        & $24^3\times32$  & 0 & 2 & 32  & 5.3   &  0.1355  & 0.16~\cite{Gockeler:2004rp} & 0.325(10) \\
D        & $12^4$          &$A\to0$& 2 & 32  & 5.6215&   0.136665 & 0.11~\cite{DellaMorte:2004bc} & - \\
\hline
\end{tabular}
\end{center}
\caption{The parameters of the  physical systems considered. A boundary field (BF) 0
means that it is vanishing, while $A$ refers to `point $A$' of Ref.~\cite{Luscher:1993gh}.}
\la{tab:param}
\end{table}

The observables we will focus on are those which typically have 
autocorrelation times much larger than that of the plaquette.
One observable (defined already in the pure gauge theory)
is $dS/d\eta$; the inverse of its expectation value defines the Schr\"odinger
functional renormalized coupling constant~\cite{Luscher:1993gh}.
The others are fermionic: the correlators $f_{\rm A}(x_0)$ and $f_{\rm P}(x_0)$
correspond to the propagation of a quark and an antiquark from a boundary to a 
point in the bulk of the lattice where the axial current or pseudoscalar density
annihilates them~\cite{Guagnelli:1999zf}. Finally we also consider $f_1$, which is 
the amplitude for boundary-to-boundary propagation through the lattice.
It serves to normalize the other correlators. In particular, 
$ Z_{\rm A} f_{\rm A}(x_0) / \sqrt{f_1}\sim \Fps  e^{-(x_0-T/2)\mps}$ holds 
far from the boundaries, where $\mps$ and $\Fps$  correspond to the pseudoscalar 
mass and decay constant~\cite{Guagnelli:1999zf}.
\subsection{HMC in the pure gauge theory\la{sec:pg}}
\begin{figure}
\vspace{-1cm}
\begin{center}
\psfig{file=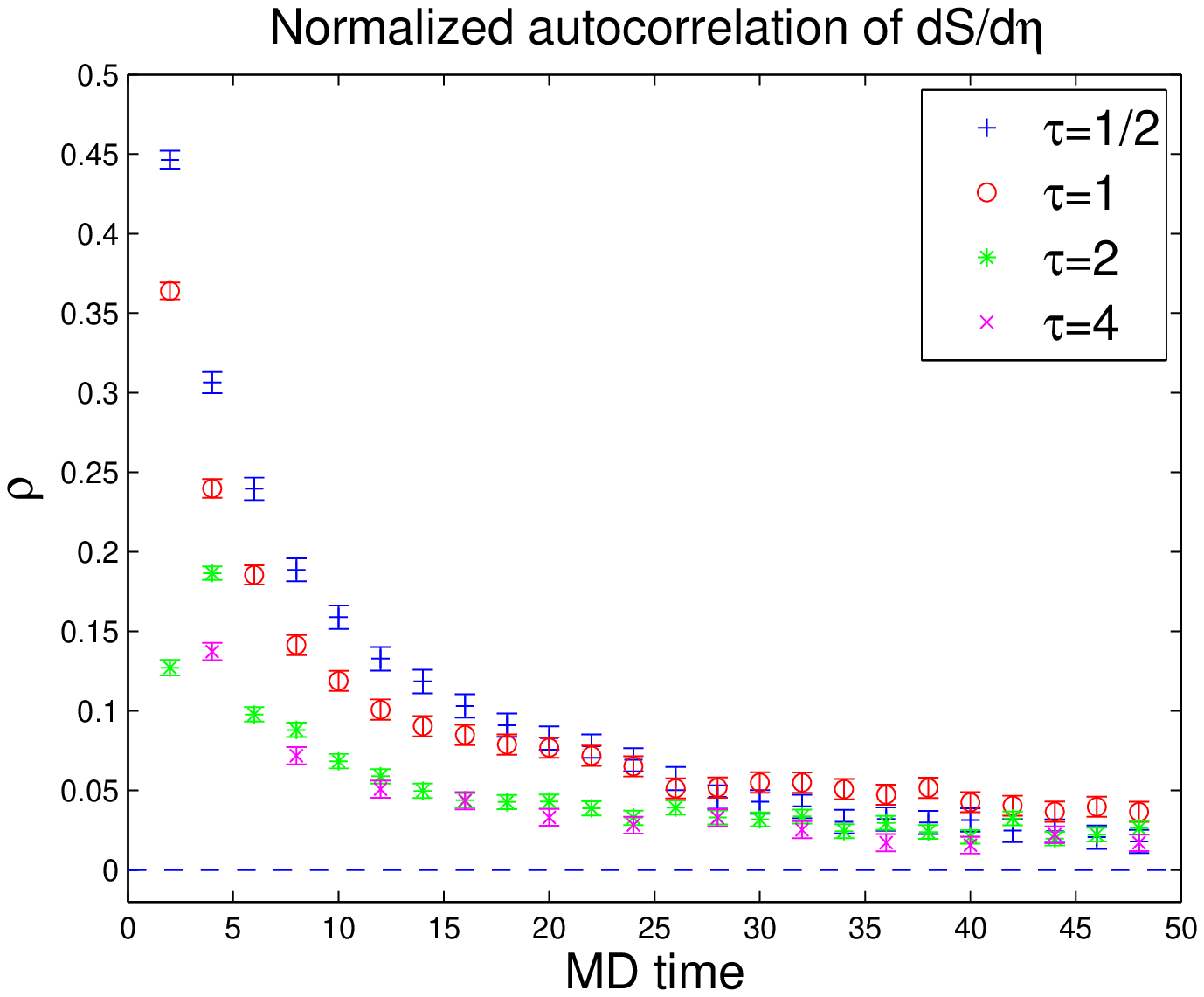,angle=0,width=9.0cm} \\
\psfig{file=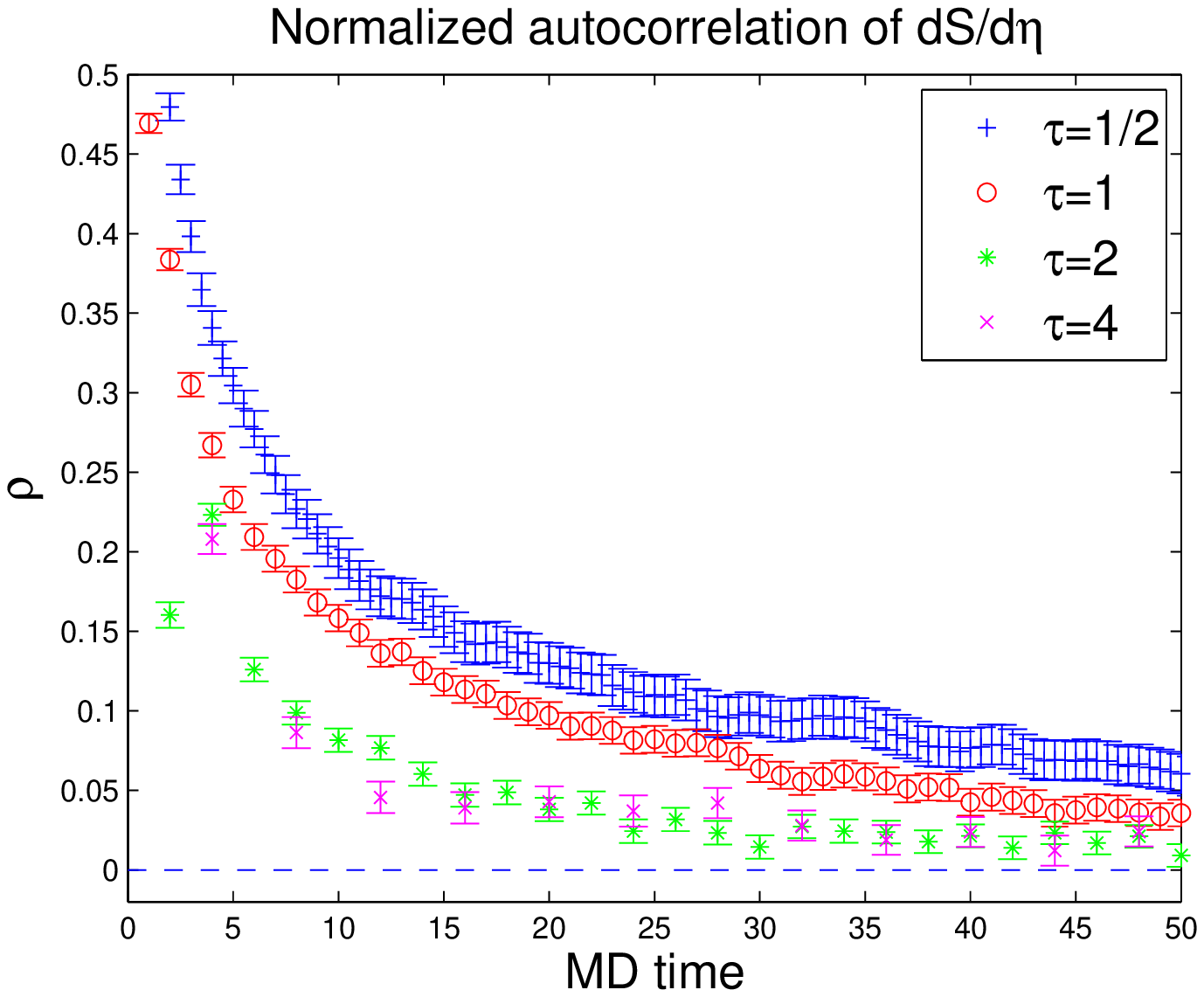,angle=0,width=9.0cm} \\
\psfig{file=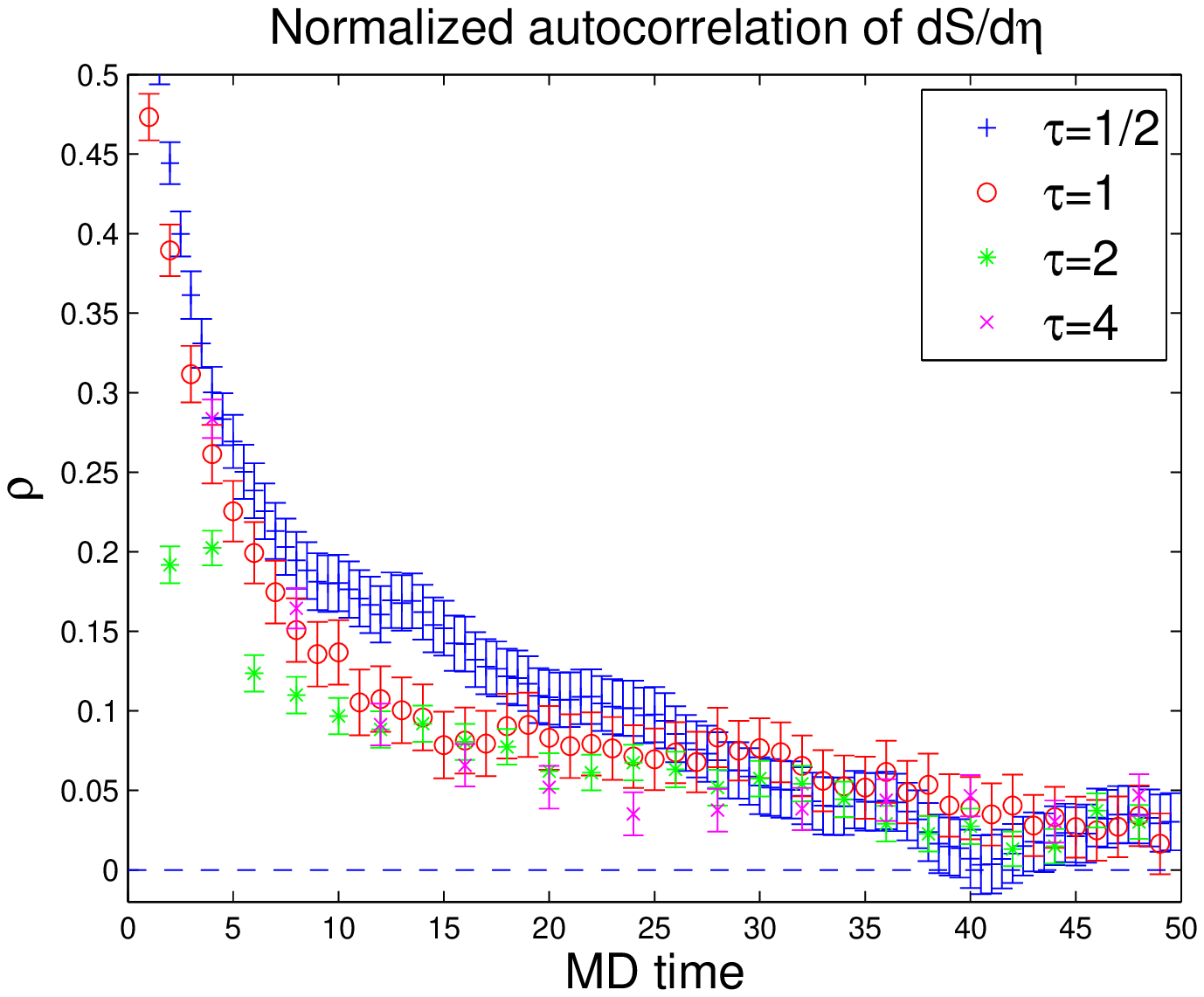,angle=0,width=9.0cm}
\end{center}
\caption{Autocorrelation function of $dS/d\eta$ in a pure
gauge HMC (system A), for different trajectory lengths (from top to bottom:
$L=8, 12$ and 16).}
\la{fig:rho[dSdeta]}
\end{figure}

\begin{figure}
\begin{center}
\psfig{file=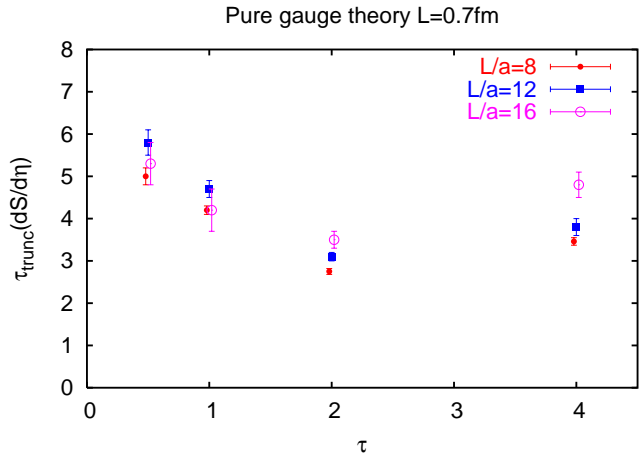,angle=0,width=9.0cm}
\end{center}
\vspace{-0.5cm}
\caption{The truncated autocorrelation time of 
$dS/d\eta$ in the pure gauge system (A), 
as a function of the trajectory length $\tau$ and 
for different lattice spacings.}
\la{fig:tauint[dSdeta]}
\end{figure}

\begin{table}
\begin{center}
\begin{tabular}{|c|c|c|c|c|}
\hline
A: pure gauge               &  $\tau=1/2$ & $\tau=1$  & $\tau=2$ & $\tau=4$ \\
\hline
\underline{$L=8$} $\tau_{\rm int}[dS/d\eta]$ &    $6.10(35)_{W=64}$ & $5.90(30)_{W=36}$  
                                    & $3.14(10)_{W=36}$  & $3.92(12)_{W=44}$\\
   $\tau_{\rm trunc}[dS/d\eta]_{W=25}$ &     5.0(2)   & 4.2(1)   & 2.75(7)   & 3.46(9)  \\
 Acceptance $[\%]$      &   97 &  97  & 97 & 96      \\
$\<\delta H^2\>/(Ld\tau)^4$ & 0.789(6)  & 0.953(6)  & 1.221(8) & 1.71(1)  \\
   $10^4\cdot\<|\delta H|_\leftrightarrow\>$ & 2.0 & 2.6 & 3.5 & 4.9        \\
\hline
\underline{$L=12$} $\tau_{\rm int}[dS/d\eta]$ & $9.5(1.0)_{W=86}$ & $6.4(5)_{W=63}$  
                               & $3.34(18)_{W=34}$  & $4.36(24)_{W=44}$ \\
 $\tau_{\rm trunc}[dS/d\eta]_{W=25}$ & 5.8(3)  & 4.7(2)   & 3.1(1)   & 3.8(2) \\
 Acceptance $[\%]$     & 94 & 93 & 93 & 91 \\
$\<\delta H^2\>/(Ld\tau)^4$ & 0.941(6) & 1.134(8) & 1.40(1) & 1.92(3)  \\
 $10^4\cdot\<|\delta H|_\leftrightarrow\>$ &5.1 & 6.4 & 8.8 & 12.2 \\
\hline
\underline{$L=16$} $\tau_{\rm int}[dS/d\eta]$ &  $6.5(1.0)_{W=52}$ & $5.8(6)_{W=52}$   
                                  & $4.2(4)_{W=40}$  & $6.0(4)_{W=68}$ \\
 $\tau_{\rm trunc}[dS/d\eta]_{W=25}$ &  5.3(5)    & 4.2(5)  & 3.5(2)  & 4.8(3) \\
 Acceptance $[\%]$  &   89  &  88  & 86  & 84  \\
$\<\delta H^2\>/(Ld\tau)^4$ & 1.08(1) & 1.31(2) & 1.58(2) &  2.10(4) \\
 $10^4\cdot\<|\delta H|_\leftrightarrow\>$ & 9.1 & 12 & 16 & 21 \\
\hline
\end{tabular}
\end{center}
\caption{The autocorrelation times (in units of MD time)
for $dS/d\eta$, the acceptance, the energy and reversibility violations
for different choices of trajectory length. $W$ is the window in MD time
over which the autocorrelation time has been accumulated. From top to bottom:
$L=8, 12$ and 16, while $La/r_0$=const. Runs in single precision.}
\la{tab:dSdeta_pg}
\end{table}

We start by investigating autocorrelation times for gluonic quantities 
in the pure-gauge HMC (system A), where we have three different lattice
spacings at constant physical parameters (see \tab\ref{tab:dSdeta_pg}). 
The acceptance rates, as well as $\<\delta H^2\>$, are given (the 
step-size is the same in all these runs); we are clearly 
in the regime of high acceptance $P_A$ where $\<\delta H^2\>\simeq 2\pi(1-P_A)^2$~\cite{MM}.
The dependence of $\<\delta H^2\>$ on $\tau$ is moderate. Therefore the dependence of $P_A$ 
is even weaker, but note that the effect of lower acceptance 
at larger $\tau$ is automatically taken into account 
in the autocorrelation  of observables.
The reversibility violations as measured by $\<|\delta H|_\leftrightarrow\>$
grow as $\sqrt{\tau}$ here, or even more slowly;
the reader is referred to the appendix for a more detailed discussion 
of $\<\delta H^2\>$ and $\<|\delta H|_\leftrightarrow\>$.
We now focus on the observable $dS/d\eta$ defined in~\cite{Luscher:1993gh}. 
We emphasize that, unlike the plaquette,
this is an observable dominated by long-distance fluctuations
and it typically has an autocorrelation time significantly 
larger than one.

The most direct way to evaluate the performance of 
different $\tau$-choices for a given observable
is to compare the corresponding normalized autocorrelation function,
$\rho(t)$. This is done on \fig\ref{fig:rho[dSdeta]}, for
trajectory lengths $1/2$, 1, 2, 4. 
There is a marked difference
between, say, $\tau=1/2$ and $\tau=2$ in favour of the latter, 
at all MD-time separations.
The choice $\tau=4$ is slightly superior to $\tau=2$ at short MD-time separation
at the largest lattice spacing, while the opposite is true at the 
finest lattice spacing.
Note in particular that at the shortest time separation,
$\rho$ is much smaller for $\tau=2$ than for $\tau\leq1$. In fact, the $\tau=2$ 
autocorrelation function shows a significant non-monotonic behaviour, 
as we have observed previously only for hybrid overrelaxation algorithms.

For a run of given MD-time length and measurement frequency, the error squared
of the observable is proportional to $1/\tau_{\rm int}$. 
When $\tau_{\rm int}$ is large compared 
to the measurement frequency, it corresponds to the area 
under the curves shown in \fig\ref{fig:rho[dSdeta]}.
In practice, a window $W$ 
must be chosen where to stop the summation. This may be chosen in a self-consistent
way which balances statistical against systematic errors. We use by default 
the prescription described in~\cite{Wolff:2003sm}. However, to rate relative
performances of algorithms, we also compute here an integrated autocorrelation
time $\tau_{\rm trunc}$ where the summation is truncated at a fixed window $W=25$. 
It turns out 
that typically $80\%$ of the true $\tau_{\rm int}$ has by then been accumulated;
the uncertainty on $\tau_{\rm trunc}$ is almost half of that on the full $\tau_{\rm int}$
and the hierarchy between the different trajectory lengths is at any rate maintained
(\tab\ref{tab:dSdeta_pg}).

\fig\ref{fig:tauint[dSdeta]} then illustrates the dependence of the truncated 
autocorrelation time on the trajectory length $\tau$.
One sees that this quantity is minimized somewhere around $\tau=2$, and this conclusion
is largely independent of the lattice spacing in the range considered.
Overall we see a variation by a factor two of $\tau_{\rm trunc}$ for 
$1/2\leq\tau\leq4$. Generally speaking this is a substantial variation
which translates directly into a corresponding speed-up of simulations
whose cost is dominated by the HMC.

\subsection{HMC in quenched QCD\la{sec:q}}
\begin{figure}
\begin{center}
\psfig{file=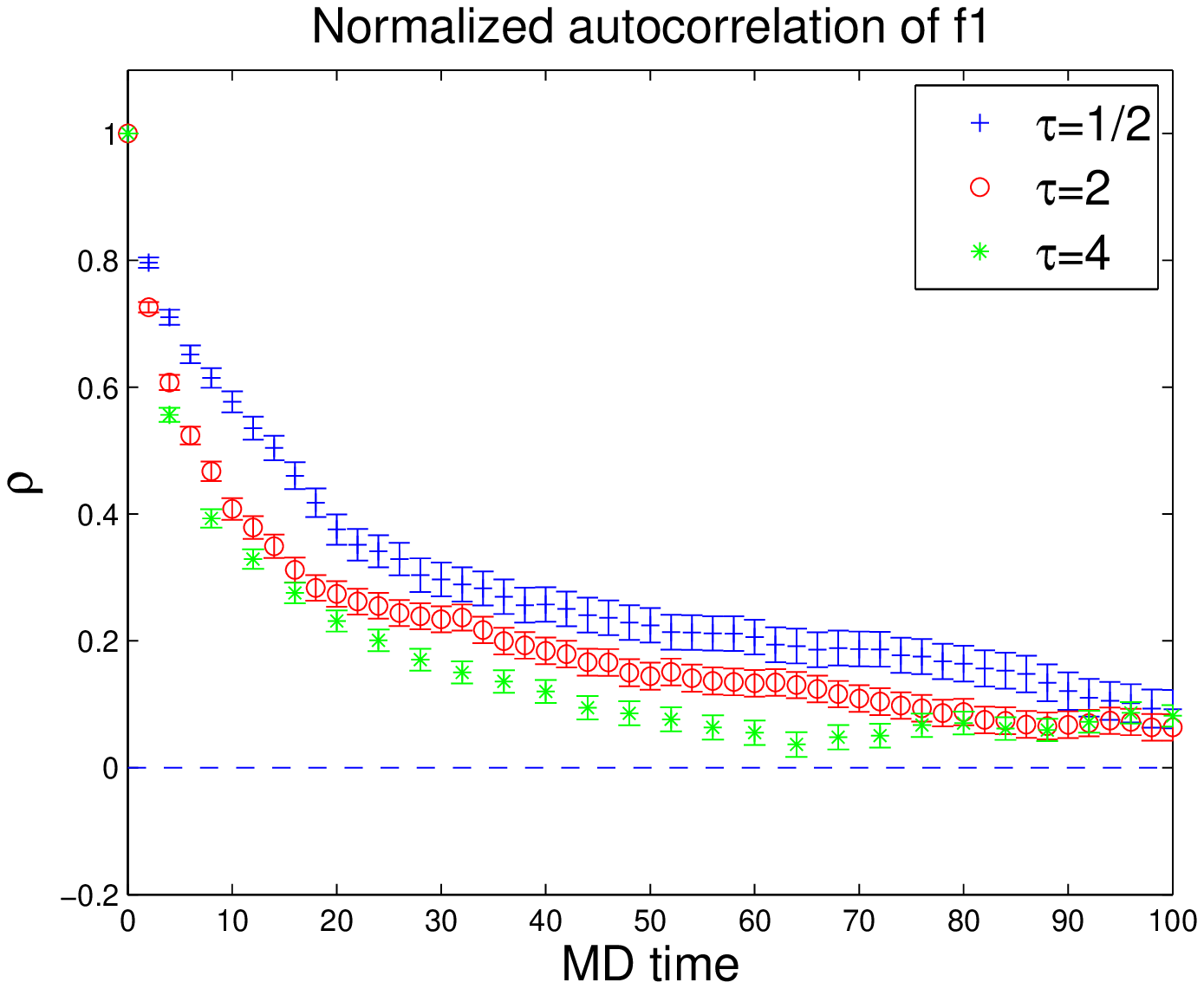,angle=0,width=7.0cm}
\psfig{file=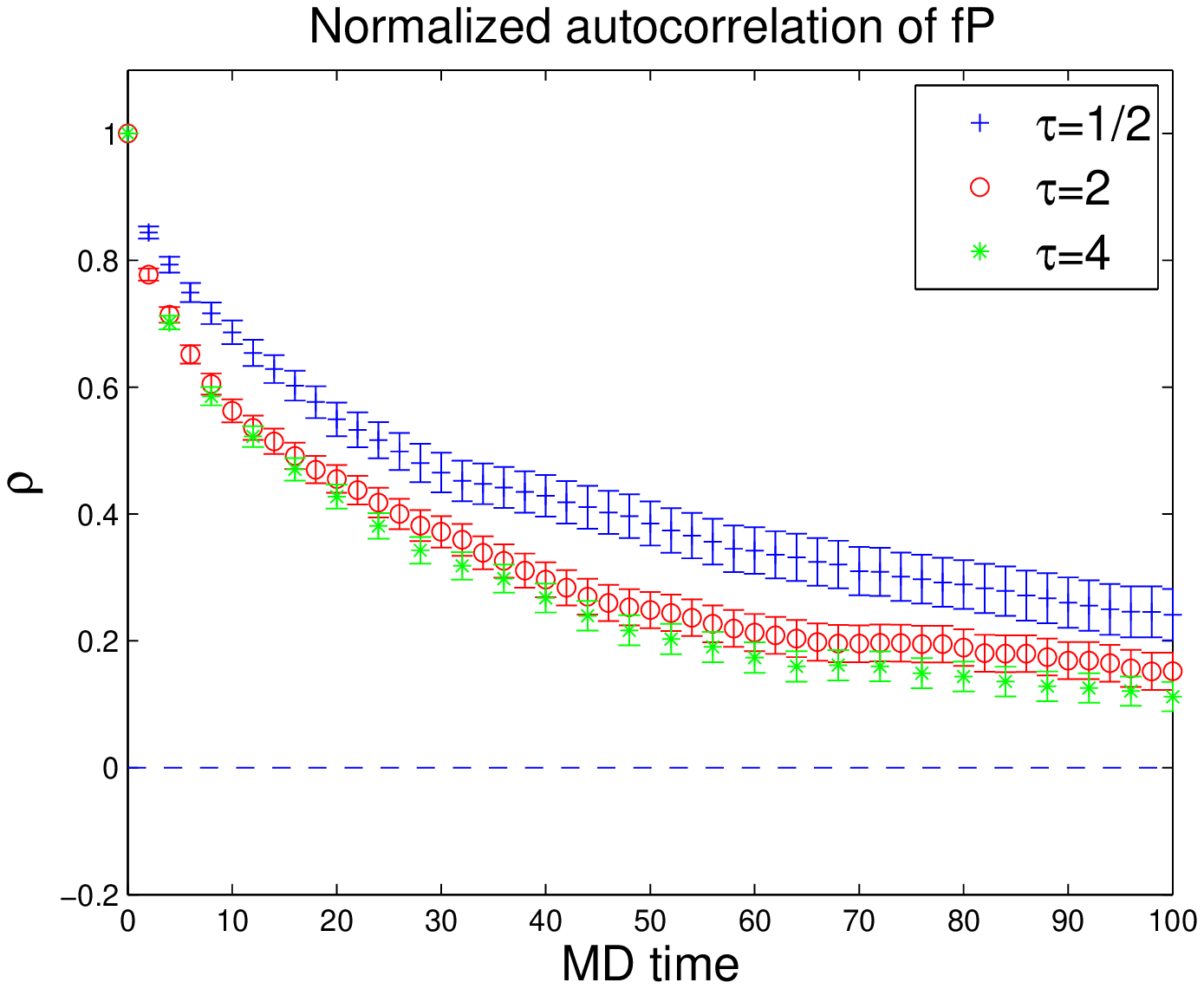,angle=0,width=7.0cm}
\end{center}
\caption{Autocorrelation function of the correlators $f_1$ and $f_{\rm P}$
in quenched QCD (system B), for different trajectory lengths.}
\la{fig:rho[f1]q}
\end{figure}

\begin{table}
\begin{center}
\begin{tabular}{|c|c|c|c|}
\hline
B: quenched, $8^3\times32$          &  $\tau=1/2$ & $\tau=2$ & $\tau=4$ \\
\hline
$\tau_{\rm int}[f_1]$ &         $40(5)_{W=270}$   &  $ 24(4)_{W=184}$ &$ 20(3)_{W=172}$  \\
$\tau_{\rm int}[f_{\rm P}]$ &          $75(20)_{W=245}$  & $44(8)_{W=316}$ &  $32(4)_{W=256}$ \\
$\tau_{\rm int}[m_{\rm PS}^{\rm eff}(T/2)]$ &      25(4) &    13.5(1.6) &     12.1(1.2)     \\
$\tau_{\rm int}[f_{\rm PS}^{\rm eff}(T/2)]$ &      33(6) &     19.0(2.6)&    14.7(1.6)      \\
Acceptance $[\%]$                      &      98    &   98       &  97         \\
$\<\delta H^2\>/(L^3Td\tau^4)$ &  0.827(5) & 1.32(1)    &1.87(3)    \\
$10^4\cdot\<|\delta H|_\leftrightarrow\>$ &   4.9   &   9.9    &   11        \\
\hline
\end{tabular}
\end{center}
\caption{The autocorrelations time (in units of MD time)
of two primary observables ($f_1$ and $f_{\rm P}$) and 
of the effective pseudoscalar mass and decay constant in the middle of the lattice, 
the acceptance, the energy and  reversibility violations
for different choices of trajectory length. Runs in single precision.}
\la{tab:Fpi_q}
\end{table}

As our next observables we now consider fermionic correlators. 
Although we are ultimately interested in dynamical, 
large-volume simulations (system C), we first investigate
the autocorrelation times in system B, which, roughly speaking, 
is a quenched version of system C with in addition the spatial extent divided by 3.
The computing time is now dominated by the measurements. The run-length
is  $t_{\rm run}\simeq32000$ in total 
(four independent lattices, called `replica', were simulated).
We investigate the range $\tau=1/2$ to $\tau=4$, within which we see hardly
any variation of the acceptance, and $\<|\delta H|_\leftrightarrow\>$
scales roughly with $\sqrt{\tau}$: \tab\ref{tab:Fpi_q} shows this, 
as well as some relevant autocorrelation times.

The comparison of autocorrelation functions for $f_1$ and $f_{\rm P}$ is done 
on \fig\ref{fig:rho[f1]q}. The advantage of $\tau=4$ and $\tau=2$  over
$\tau=1/2$ is both substantial and statistically significant. This is reflected
in a factor of about two in the integrated autocorrelation times.
This gain is at least as marked in physical quantities such as 
the effective pseudoscalar mass and decay constant.

\subsection{Mass-preconditioned HMC in $\Nfii$ QCD\la{sec:nf2}}
\begin{figure}
\begin{center}
\psfig{file=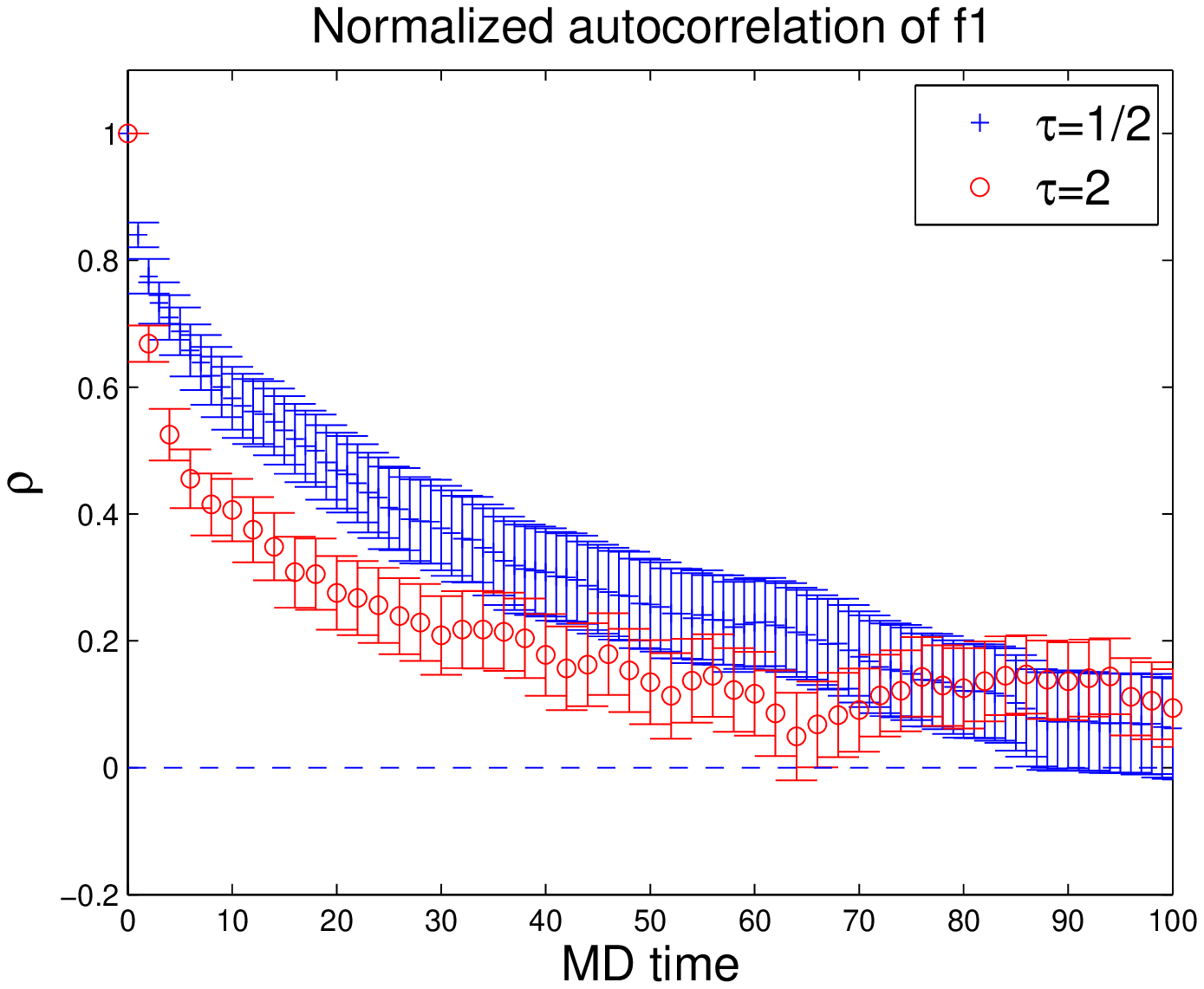,angle=0,width=7.0cm}
\psfig{file=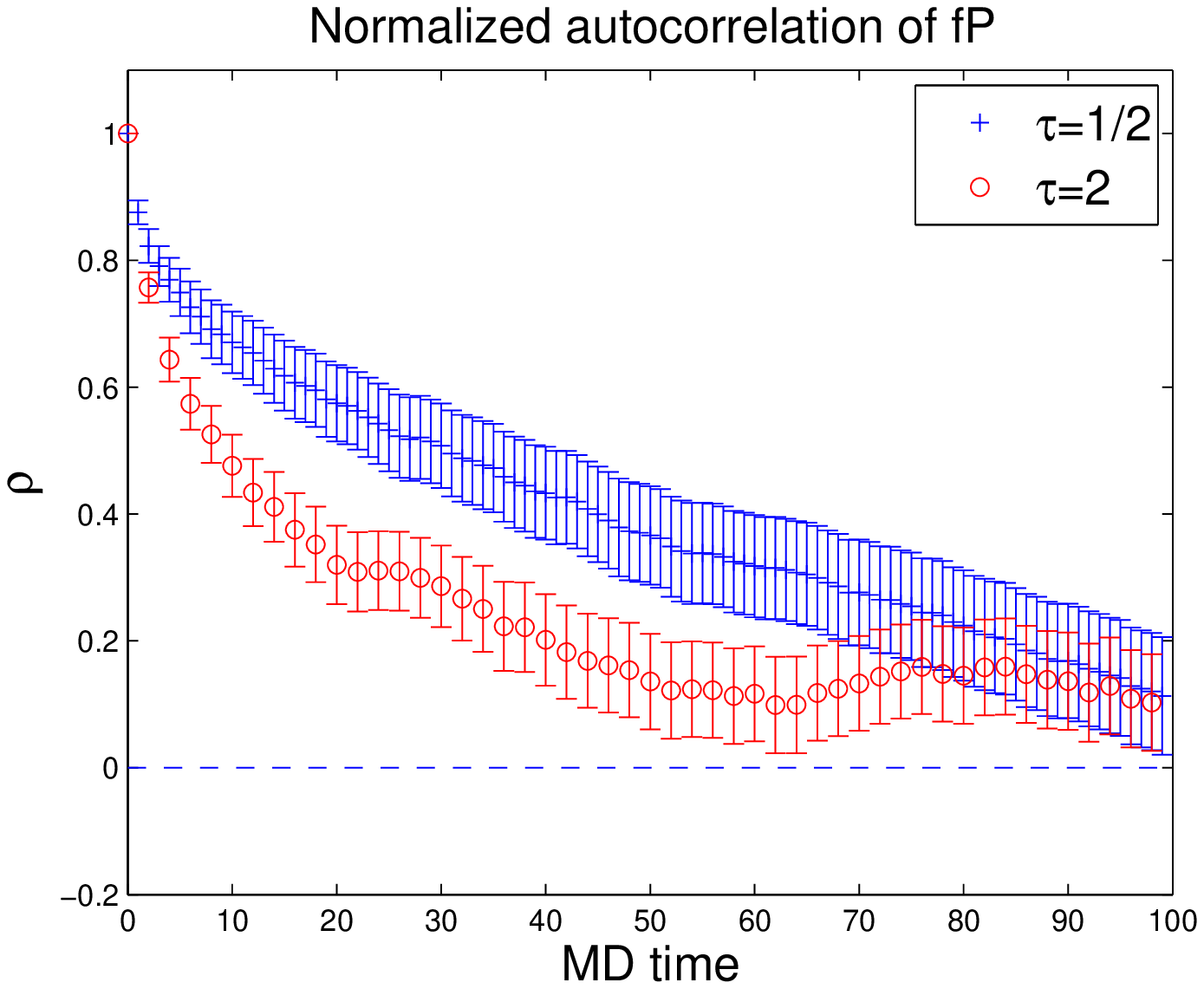,angle=0,width=7.0cm}
\end{center}
\caption{Autocorrelation function of the correlators $f_1$ and $f_{\rm P}$
in $\Nfii$ QCD (system C), for different trajectory lengths.}
\la{fig:rho[f1]nf2}
\end{figure}

\begin{table}
\begin{center}
\begin{tabular}{|c|c|c|}
\hline
C: $\Nfii,~\kappa=0.1355$ &              $\tau=1/2$ & $\tau=2$ \\
\hline
$\tau_{\rm int}[f_1]$ &       $30(15)_{W=145}$      &   $24(8)_{W=134}$   \\
$\tau_{\rm trunc}[f_1]_{W=50}$ & 23(5)          & 15(3)   \\
$\tau_{\rm int}[f_{\rm P}]$ &        $45(20)_{W=185}$      &  $26(10)_{W=134}$   \\
$\tau_{\rm trunc}[f_{\rm P}]_{W=50}$ &  28(7)      &  18(4)   \\
$\tau_{\rm int}[m_{\rm PS}^{\rm eff}(T/2)]$ &  9.5(2.5)   &    6.8(1.5) \\
$\tau_{\rm int}[f_{\rm PS}^{\rm eff}(T/2)]$ &  14(4)      &    4.2(8)   \\
Acceptance $[\%]$                      &   90        &     91       \\
$\<\delta H^2\>/(L^3Td\tau^4)$                       & $0.147(4)^*$ & 0.164(6)  \\ 
$10^4\cdot\<|\delta H|_\leftrightarrow\>$ & 1.0        &     2.5       \\
\hline
\end{tabular}
\end{center}
\caption{The autocorrelations time (in units of MD time)
of two primary observables ($f_1$ and $f_{\rm P}$) and 
of the effective pseudoscalar mass and decay constant in the middle of the lattice,
the acceptance, the energy and reversibility violations
for different choices of trajectory length. The value of $\<\delta H^2\>/(L^3Td\tau^4)$ 
marked by a $^*$ was obtained for one of the replica; the other had one spike of 
$\delta H \simeq2000$. Runs in double precision.}
\la{tab:Fpi_nf2}
\end{table}

We now come to $\Nfii$ dynamical simulations in a spatial volume of 
$(2{\rm fm})^3$, and consider the same fermionic correlators 
as in the previous section. 
The quark mass is around  $m_s$, the strange quark mass,
and we have two values of $\tau$, $1/2$ and 2.
The total statistics in each case is about $t_{\rm run}=4000$ (2 replica were simulated).

Note that the acceptance is practically the same in both runs.
The  autocorrelation functions of $f_1$ and $f_{\rm P}$ are compared 
on \fig\ref{fig:rho[f1]nf2}. Naturally the error bars are now larger;
nonetheless a statistically significant advantage of the run 
at $\tau=2$ is seen at MD-time separations up to about 30. 
This is confirmed when one looks at the truncated 
autocorrelation time with a window of $W=50$.
The autocorrelation function of $f_1$ is in fact surprisingly similar 
with that in the quenched simulation, \fig\ref{fig:rho[f1]q}.
The resulting autocorrelation times are given in \tab\ref{tab:Fpi_nf2}.
In spite of large uncertainties, there is a significant reduction of 
$\tau_{\rm int}[f_{\rm PS}^{\rm eff}(T/2)]$ when going from $\tau=1/2$ to 
$\tau=2$. 
\subsubsection{Thermalization with light quarks}
Since we observe 
that autocorrelation times of long-distance observables 
are reduced by the use of longer trajectories,
we also expect their thermalization to be accelerated by the latter.
Consider the thermalization of system D (see \tab\ref{tab:param}).
The exercise here is to 
\begin{enumerate}
\item start from the ensemble with non-zero boundary field
(`point A' of~\cite{Luscher:1993gh});
\item revert to homogeneous Dirichlet boundary conditions, and hence
vanishing background field~\cite{Luscher:1993gh};
\item let the system rethermalize under the HMC evolution, 
for two different choices of trajectory length;
\item track how fast quantities whose expectation values vanish
by symmetry in the homogeneous Dirichlet case relax to zero.
\end{enumerate}

\fig\ref{fig:therm}
shows the relaxation of the observables $dS/d\eta$ and $f_{\rm P}-f_{\rm P}'$. 
The latter is the asymmetry between the correlator emanating from one 
boundary and the other. A data point at time $t$
shows the value of the observable
averaged over 16 independent replica (corresponding to independent 
starting configurations), and averaged over the MD 
time interval $]t-6,t+6]$. The $dS/d\eta$ measurements  
were done after every trajectory and  $f_{\rm P}-f_{\rm P}'$ every other trajectory.

\begin{figure}
\begin{center}
\psfig{file=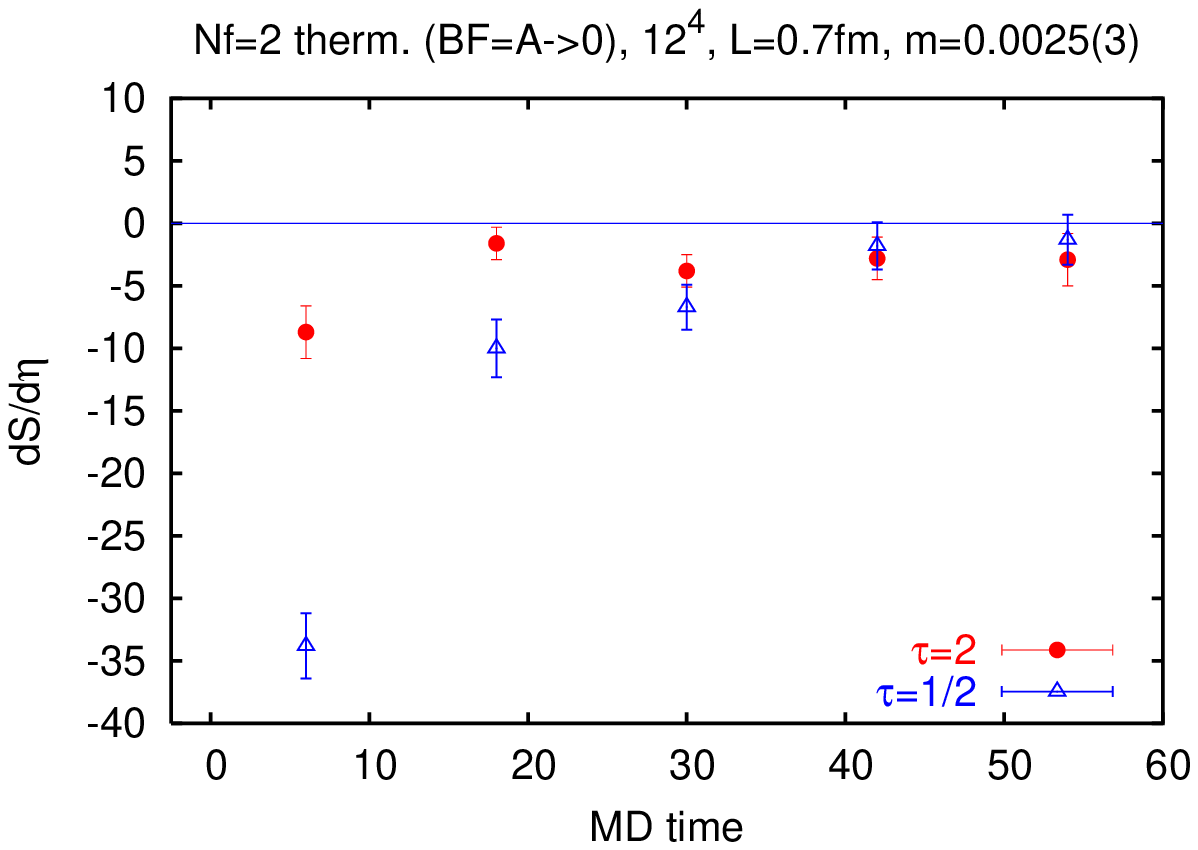,angle=0,width=7.0cm}
\psfig{file=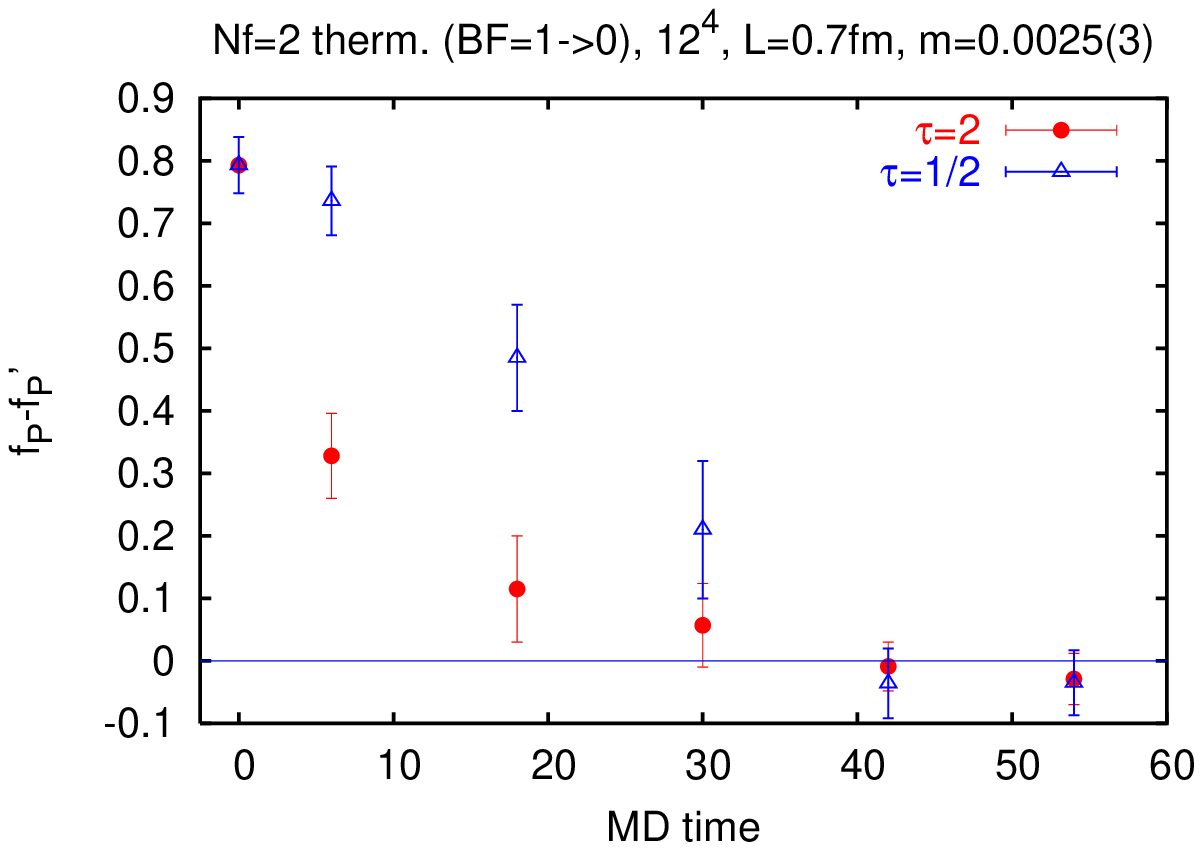,angle=0,width=7.0cm}
\end{center}
\vspace{-0.5cm}
\caption{Thermalization of two long-distance observables ($dS/d\eta$
and $f_{\rm P}-f_{\rm P}'$) having zero as expectation value. The 
starting configurations are taken from an ensemble where 
these expectation values are non-zero (system D).}
\la{fig:therm}
\end{figure}

The thermalization takes place faster with the choice of 
trajectory length $\tau=2$, at least in the relatively early
stages. In difficult simulations the time-consuming
part is presumably the tail of the thermalization, but 
it would require very large statistics to demonstrate an
algorithmic advantage in that regime.

\section{Conclusion\la{sec:conc}}
We have investigated the dependence of autocorrelation times
on the HMC trajectory length, focusing on long-distance observables, 
in a variety of different physical situations. 
We find this dependence to be substantial (a factor around 2) 
and statistically significant.
The reduced correlation of successive measurements, 
done at fixed intervals of molecular dynamics time,
is most clearly seen in the autocorrelation functions themselves.
This means that a reasonable tuning of the trajectory
length may save a factor of about 2 in computing time. 

The optimal choice of $\tau$ is observable dependent.
However we have observed that trajectories longer than the ones
conventionally used provide an advantage in computing 
standard physical quantities such as the pseudoscalar 
mass and decay constant. It will be interesting to see
whether this conclusion also holds for HMC algorithms 
that use a different preconditioning of the pseudofermion 
action, for a different number of flavours, etc.

Naturally, other criteria are relevant in the final choice of trajectory length;  
the issues of stability and reversibility violations have been addressed in the appendix
(see also Ref.~\cite{DellaMorte:2003jj,Liu:1997fs,Joo:2000dh} to name a few).
We are currently performing a simulation at a quark mass of $m_s/2$ with $\tau=2$
that shows good stability and controlled reversibility violations.

\section*{Acknowledgements}
We thank DESY/NIC for computing resources on the APE machines
and the computer team for support, in particular to run on the new
apeNEXT systems.
This project is part of ongoing algorithmic development within
the SFB Transregio 9 ``Computational Particle Physics'' programme.
R.S. thanks the Yukawa Institute for Theoretical Physics at Kyoto 
University; discussions during the YITP workshop 
\emph{Actions and symmetries in lattice gauge theory}
YITP-W-05-25 were useful to this work. 

\appendix
\section*{Appendix: Stability and reversibility violations}
\begin{figure}
\begin{center}
\psfig{file=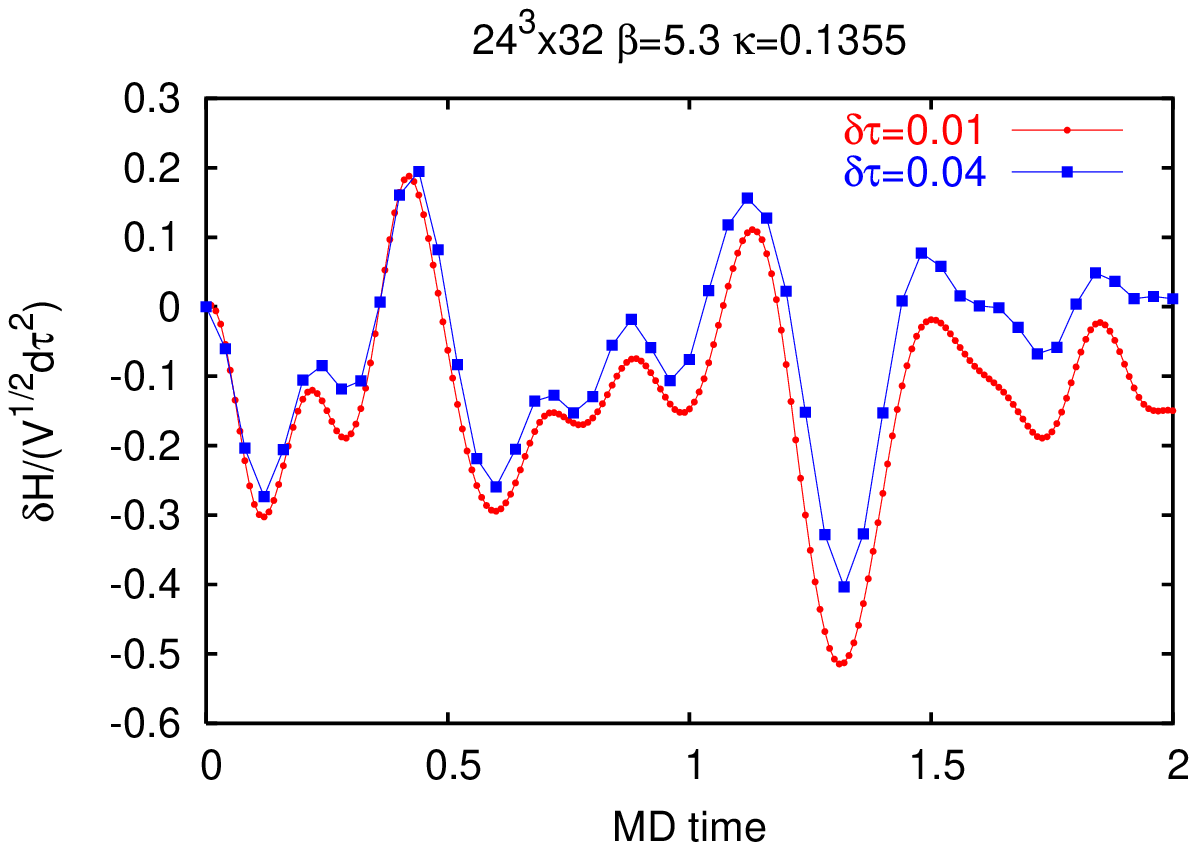,angle=0,width=7.0cm}
\psfig{file=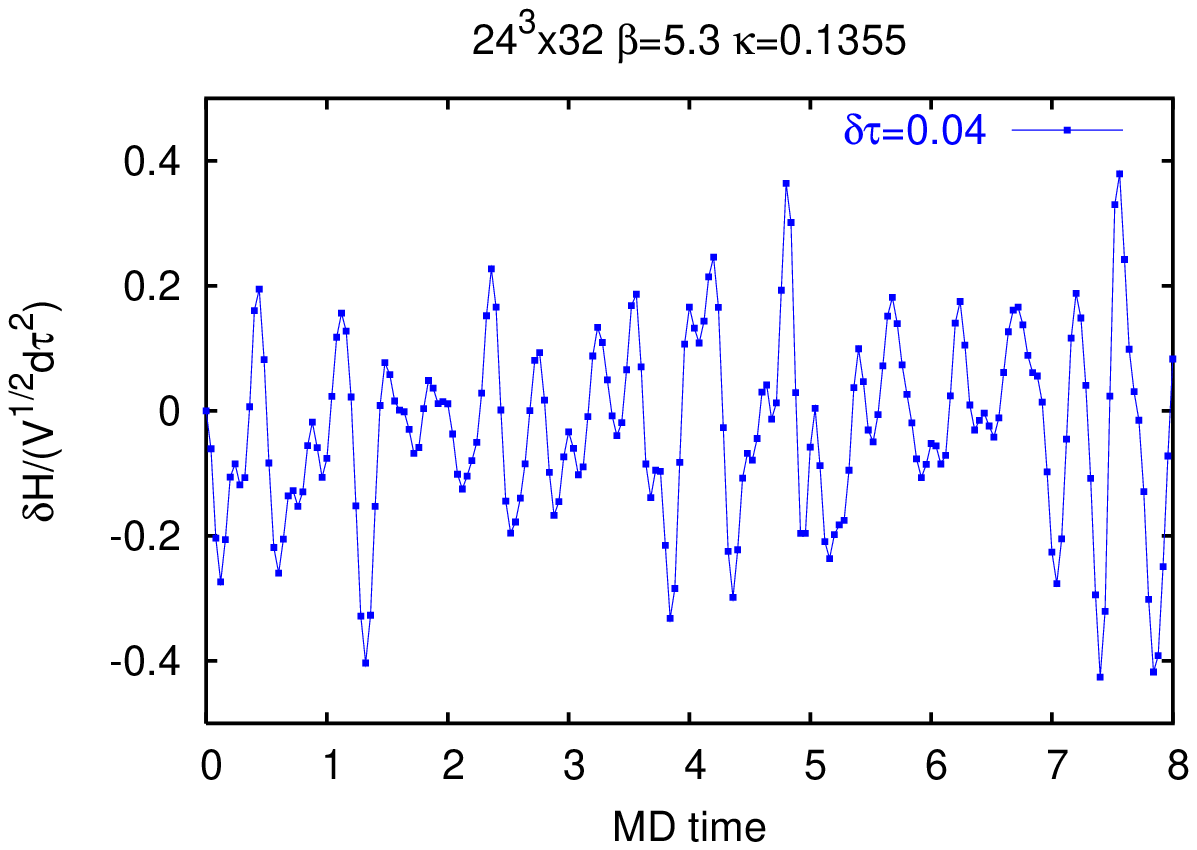,angle=0,width=7.0cm}
\end{center}
\vspace{-0.5cm}
\caption{Variation of the Hamiltonian along one MD trajectory on lattice C,
in units of $\sqrt{V}d\tau^2$, where $V$ is the number of lattice points.}
\la{fig:dH}
\end{figure}

In this appendix we discuss some important issues that one might worry
about if one increases the length of MD trajectories. Most of what follows
is not new but we find it useful to gather here the relevant points.

\subsubsection*{Energy violations along a trajectory}
\fig\ref{fig:dH} shows the variation of the Hamiltonian along one  
MD trajectory on lattice C, for two different step sizes 
(here the leap-frog integrator was used 
with a ratio of 5 between the time-steps of the two pseudo-fermion forces
\cite{AliKhan:2003br}). The right plot illustrates the fact that the fluctuations 
of $\delta H(t)$ around zero do not grow very fast with MD time $t$.

Although a symplectic integrator such as leap-frog has discretization errors,
there is a Hamiltonian which differs by O$(d\tau^2)$ terms from the MD Hamiltonian, and
which is \emph{exactly} conserved by the MD evolution~\cite{Gupta:1988js,kennedy}:
\be
H_{\rm o} = H(t) + d\tau^2 h_1(t) + d\tau^4 h_2(t) +\dots
\ee
Hence, for a given start configuration and momenta,
\be
- \frac{\delta H(t)}{d\tau^2} =  \delta h_1(t) + d\tau^2 \delta h_2(t) + \dots 
\ee
In words, the curve $\delta H(t)/d\tau^2$ is independent of $d\tau$, up to 
O$(d\tau^2)$  corrections.
This is clearly what is seen on the left plot of \fig\ref{fig:dH}.
Note that since in equilibrium 
$\<\delta H \>\simeq 1/2\<\delta H^2\> = {\rm O}(d\tau^4) >0$ holds,
we have $\<\delta h_1(t)\>=0$ and $\<\delta h_2(t)\><0$ for any fixed $t$.

\subsubsection*{Scaling of energy violations}
The scaling law $\<\delta H^2\>\sim Vd\tau^4$ 
was proposed in~\cite{Gupta:1988js,Creutz:1988wv}.
Indeed, since $\delta H$ is O$(d\tau^2)$ for one trajectory, 
$\delta H^2$ is O$(d\tau^4)$ and so is its average. 
As to the volume dependence, we confirm that $\<\delta H^2\>$ depends mainly 
on the number of lattice points $V$~\cite{Gupta:1990ka}, 
while the dependence on the coupling $\beta$ is very weak\footnote{
Additional $\beta=6.086$ simulations at $L=12$ and 16 in system A with $\tau=2$
give $\<\delta H^2\>/(Ld\tau)^4=1.31(2)$ and 1.36(6) respectively.}.
The values of $\<\delta H^2\>$ in \tab\ref{tab:dSdeta_pg} are consistent
with $\<\delta H^2\>\propto V^{\eta(\tau)}$, where $\eta\in[1.07,1.11]$.

The dependence of $\<\delta H^2\>$ on the trajectory length $\tau$ is very moderate
but follows no obvious formula (see \tab\ref{tab:dSdeta_pg}). 
Nevertheless, the trend is reasonably well described by $\tau^\alpha$, $0.3<\alpha<0.4$.

\subsubsection*{Stability}
Of course the statements made in the previous paragraph
hold for exact arithmetics, 
and in practice it must be checked empirically at what trajectory length
rounding errors introduce instabilities in the integration~\cite{Joo:2000dh}.
The other source of instability can be the occurrence of an
exceptionally large force, which spoils the expansion in $d\tau$.
In the simulation C (see \tab\ref{tab:param}), $|\delta H|$ was 
bounded by 1 in the $\tau=2$ run, while one replicum experienced 
one spike of  $\delta H\simeq 2000$ in the $\tau=1/2$ run
(the simulation then continued normally).

\subsubsection*{Reversibility violations}
Reversibility violations in our simulations 
come from rounding errors, and from the non-zero residuals of the inversions
in the MD evolution\footnote{The solver starts from the  vector zero  
in our implementation of the molecular dynamics.}. These effects 
of course accumulate with the length of the trajectory. However
\tab\ref{tab:dSdeta_pg}, \ref{tab:Fpi_q} and \ref{tab:Fpi_nf2} show 
that the increase of $\<|\delta H_{\leftrightarrow}|\>$ with $\tau$
in simulations  where $|\delta H|$ is bounded by one
and is typically much smaller, is rather slow,  roughly
like $\sim\sqrt{\tau}$. Therefore it should not present 
a serious problem. In the trajectory of \fig\ref{fig:dH} with $d\tau=0.04$,
$10^4\cdot\<|\delta H|_\leftrightarrow\>$ is 0.06, 0.3, 1.4 and 1.4 for $\tau=0.08$, 
0.8, 1.6 and 8 respectively. 

How a small reversibility violation influences the ensemble generated 
is not known (to us); it may lead to an incorrect sampling.
The influence on expectation values of observables is even harder to pin down. 
The study~\cite{DellaMorte:2003jj} showed that 
even with $\<|\delta H_{\leftrightarrow}|\>=0.01$ one does not see an 
effect in a number of observables computed at the sub-percent level.


\end{document}